\documentclass[aps,prl,twocolumn,showpacs]{revtex4}

\usepackage{graphicx}
\usepackage{amsfonts}
\usepackage{amssymb}
\usepackage{amsmath}
\usepackage{mathrsfs}
\usepackage{braket}
\usepackage{color}

\begin{document}

\title{Nearly flat Andreev bound states in superconductor-topological insulator hybrid structures}
\author{Mahmoud Lababidi}
\author{Erhai Zhao}
\affiliation{School of Physics, Astronomy, and Computational Sciences, George Mason University, Fairfax, VA 22030}
\begin{abstract}
Exotic excitations arise at the interface between a three-dimensional topological insulator (TI) and superconductors.
For example, Majorana fermions with a linear dispersion, $E\sim k$, exist in a short $\pi$ Josephson junction on the TI surface.
We show that in these systems, the Andreev bound states spectrum becomes nearly flat at zero energy when the chemical potential is sufficiently away from the Dirac point.
The flat dispersion is well approximated by $E\sim k^N$, where $N$ scales with the chemical potential. 
Similar evolution from linear to flat dispersion also occurs for the subgap spectrum of a  
periodic superconducting proximity structure, such as a TI surface in contact with a stripe superconductor.

\end{abstract}
\maketitle

Moving at ``the speed of light", $v_F$,
massless Dirac electrons on the surface of a three-dimensional $Z_2$ topological insulator (TI) can not be localized by scattering from nonmagnetic impurities \cite{hasan_colloquium:_2010,qi_topological_2011}, nor can they be
easily confined by electrostatic potentials due to Klein tunneling \cite{Katsnelson:2006fk}. Proximity coupling to 
ferromagnetic or superconducting order can however 
open up a gap in the spectrum, thus rendering excitations massive \cite{hasan_colloquium:_2010,qi_topological_2011}.
 An intriguing
possibility is to engineer new {\it massless} excitations by confining and coherently mixing Dirac electrons 
and holes using two or more superconductors with definite phase difference \cite{fu_superconducting_2008}.
For example, Fu and Kane showed that a Josephson junction on the surface of a TI with a phase bias of $\pi$ is a one-dimensional quantum wire for Majorana fermions, 
which can be further manipulated by using tri-junctions \cite{fu_superconducting_2008}. Signatures of Majorana fermions in such structures have been reported in recent 
experiments \cite{williams_unconventional_2012,moore_extraordinary_2012}.

In this Letter, we demonstrate a drastically different regime for the same, albeit slightly more general, Josephson structures considered by Fu and Kane.
This regime features massless zero energy excitations that are almost dispersionless, i.e. with vanishing group velocity $(\partial E/\partial k=0)$.
We elucidate the scattering kinematics behind the nearly flat dispersion at zero energy using simple models, and verify the results with self-consistent calculations.
We find it striking that in such simple structures, which are now available in experiments, the low energy excitation can be easily tuned all the way from $E\sim k$ to $E\sim k^N$, where $N$ is large, by increasing the chemical potential. 
By extending such junctions into a class of {\it periodic} superconductor-TI proximity structures, we further show that these states become a flat band near zero energy.

\begin{figure}
\center
\includegraphics[width=3.4in]{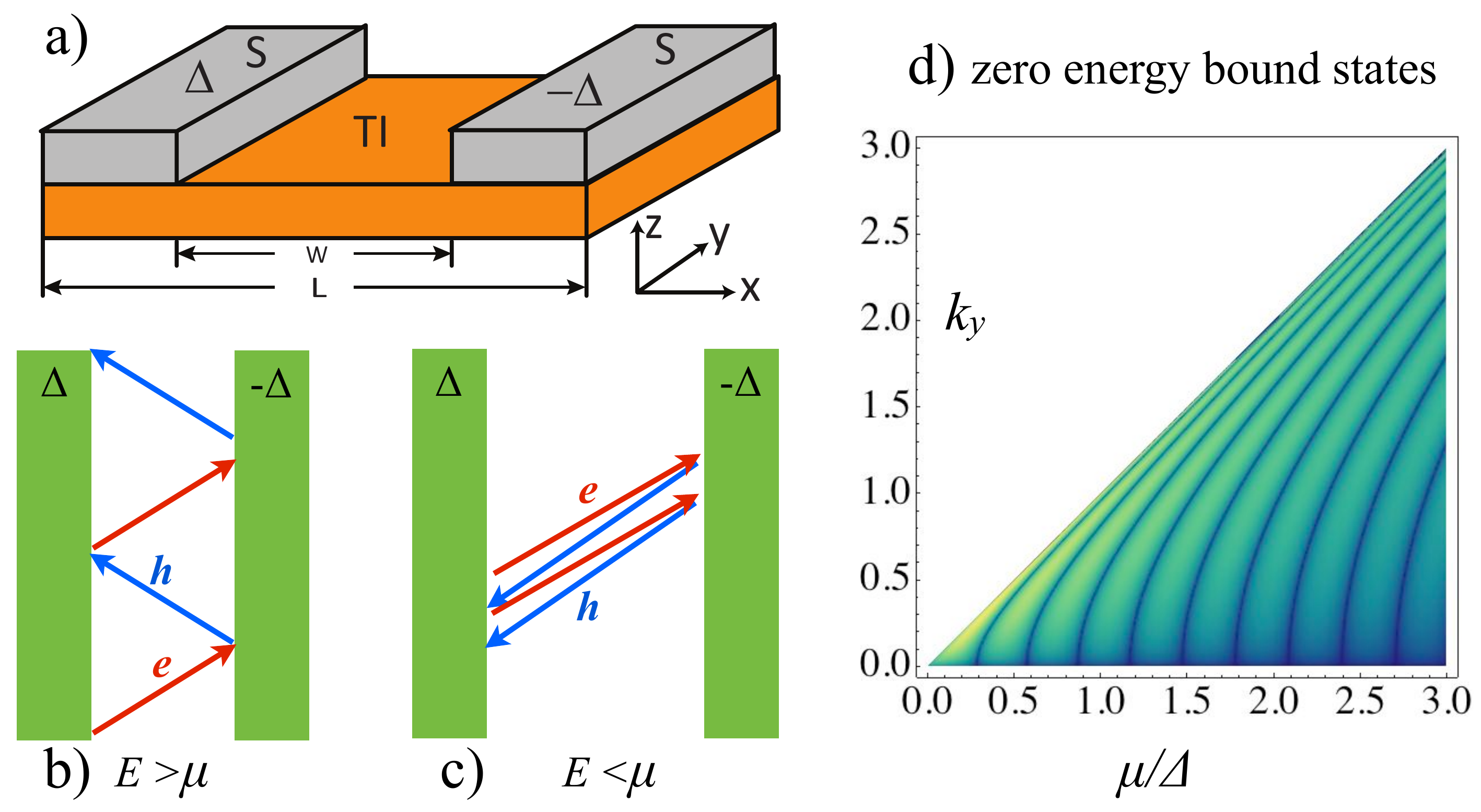}
\caption{(color online) a) Schematic of a Josephson junction on the surface of a topological insulator (TI).
The two superconducting leads (S) have a phase difference $\pi$. $\Delta$ is the superconducting gap, and 
$w$ is the junction width (not to scale).
b) Specular Andreev reflection in the regime $E>\mu$. c) Retro-reflection for $E<\mu$.
d) Dark lines show the $(k_y,\mu)$ values for the zero energy Andreev bound states for $w=10\hbar v_F/\Delta$ 
and $L\rightarrow \infty$.
$k_y$ is in unit of $\Delta/\hbar v_F$.
}\label{jjsetup}
\end{figure}

The Josephson junction is schematically shown in Fig.~\ref{jjsetup}a). Two $s$-wave superconductors are patterned
on the TI surface. Due to the proximity effect, the S-TI interface becomes a 2D superconductor (S).
The S-TI-S junction can be well described by the following Bogoliubov-Dirac Hamiltonian introduced in Ref. \cite{fu_superconducting_2008},
\begin{equation}
\mathcal{H}=\hbar v_F(\sigma_x k_y +i \tau_z\sigma_y \partial_x) + \tau_z \mu(x) + \tau_y \sigma_y \Delta(x).
\label{ham}
\end{equation}
Here $\tau_i$ ($\sigma_i$) are the Pauli matrices in the particle-hole (spin) space. 
The system is translationally invariant in the $y$ direction, and $k_y$ is the momentum along $y$. 
In the TI region of 
length $w$, the superconducting order parameter $\Delta(x)$=0, while it is constant $\Delta$ deep into the superconductor. 
The chemical potential $\mu$ can be tuned by applying a gate voltage.
In general, its value can differ in the TI and S region, but for simplicity,
we assume it is uniform in all regions. 
Also, we will focus on the case of phase difference of $\pi$ across the junction.

We first give a heuristic argument for the existence of two regimes.
A Dirac electron in the TI region incident on S will be Andreev reflected into a hole if its energy is below the superconducting gap ($E<\Delta$). 
In the context of graphene \cite{PhysRevLett.97.067007,RevModPhys.80.1337}, 
Beenakker pointed out that in addition to the familiar Andreev retro-reflection where the reflected hole has a group velocity opposite to the incident electron when $E<\mu$, 
there is also the case of specular Andreev reflection where the reflected hole's group velocity is in the specular direction for $E>\mu$. Typical scattering trajectories in these two regimes are contrasted in Fig.~\ref{jjsetup}b) and \ref{jjsetup}c).
For $\mu=0$ as considered in Ref.~ \cite{fu_superconducting_2008}, 
the Majorana fermion excitation with linear dispersion is associated with the specular Andreev reflections in Fig.~\ref{jjsetup}b). 
For large $\mu$, as in the case of as grown Bi$_2$Se$_3$ crystals, one expects very different behaviors at low energies. 
For the $E<\mu$ case, it can be shown analytically that the phase of the retro-reflected hole is equal to the incident angle of an incoming electron at zero energy, $\theta=\arcsin(\hbar v_F k_y/\mu)$. This is unique to TIs because the wavefunction of a Dirac electron [or hole], $(1,\pm e^{i\theta},0,0) [(0,0,1,\pm e^{i\theta})]$, 
is determined by the angle $\theta$, or $k_y$. The resultant hole incident on the opposite S with phase of $\pi$ retro-reflects into an electron. This electron has exactly the same phase as it started with, thus forming an Andreev bound state.

The remaining key question is whether there will be any states at or near 
zero energy when $\mu$ is finite. We can answer the question by solving 
Eq.~(\ref{ham}) for an idealized, step function profile of $\Delta(x)$,
\begin{equation}
\Delta(x) = \Delta[\theta(-x)-\theta(x-w)].
\end{equation}
The dark lines in Fig.~\ref{jjsetup}d) shows the zero energy solution in the 
$(\mu,k_y)$ plane, with fixed $\Delta$ and the junction length $w=10\hbar v_F/\Delta$. 
In general, there exist multiple zero energy bound states
at discrete $k_y$ values $\{k_y^i\}$ for finite $\mu$. 
For increasing $\mu$ and  $w$,
these solutions become increasingly close-packed.
This nontrivial result has important implications for experiments. 
The Majorana quantum wire is only ideal in the limit of $\mu,w\rightarrow 0$.
As $\mu$ is tuned away from the Dirac point, the single zero energy state at $k=0$ will 
be replaced by multiple zero energy solutions
along the $k_y$ axis, and eventually a nearly flat dispersion at zero energy.

To unambiguously establish this claim, we solve the differential
equation $\mathcal{H}(x,k_y)\psi(x,k_y)=E \psi(x,k_y)$ numerically 
for a finite size system, $x\in [0,L]$ as shown in Fig.~\ref{jjsetup}a),
with open boundary conditions at $x=0,L$ \cite{footnote1}. Here the quasiparticle wave function
$\psi=\left ( { u} _{\uparrow},  { u}_{\downarrow},  { v}_{\uparrow}, { v}_{\downarrow} \right )^T$,
with the label $(x,k_y)$ omitted.
To fully describe the proximity
effect including the induced superconducting correlations in the TI region
and the suppression of superconductivity near the TI-S boundary, we 
determine the order parameter profile $\Delta(x)$ self-consistently
through the gap equation
\begin{equation}
\Delta(x)= g(x)\sum_{\epsilon_n<\omega_D}\int d k_y u_{n,\uparrow}(x,k_y) v_{n,\downarrow}^\ast(x,k_y).
\end{equation}
Here $n$ labels the eigenstates with energy $\epsilon_n$, $g$ is the effective
attractive interaction, and $\omega_D$ is the Debye frequency. We assume $g$
is zero in the TI region and constant inside S. We expand $\psi(x,k_y)$ and 
$\Delta(x)$ in Fourier series and convert the differential equation into 
an algebraic equation~\cite{halterman_characteristic_2011,PhysRevB.83.184511}. Starting with
an initial guess of $\Delta(x)$ which features phase difference 
$\pi$, the iterative procedure is repeated until desired convergence is achieved.
Note that the phase difference $\pi$ is self-maintained throughout and not fixed by hand after every iteration.
Then, the local spectral function,
\begin{equation}
A_\sigma(E,k_y,x)=\sum_n \delta(E-\epsilon_n)|u_{n\sigma}(x,k_y)|^2,
\end{equation} 
and the local density of states (LDOS),
\begin{equation}
N(E,x)=\int dk_y\sum_{n,\sigma} \delta(E-\epsilon_n)|u_{n\sigma}(x,k_y)|^2,
\end{equation} 
can be computed for $\sigma=\uparrow,\downarrow$. The calculation is checked to
reproduce known results, e.g., the linearly dispersing Majorana spectrum
at $\mu=0$ predicted in Ref. \cite{fu_superconducting_2008}.

\begin{figure}[]
\includegraphics[width=3.4in]{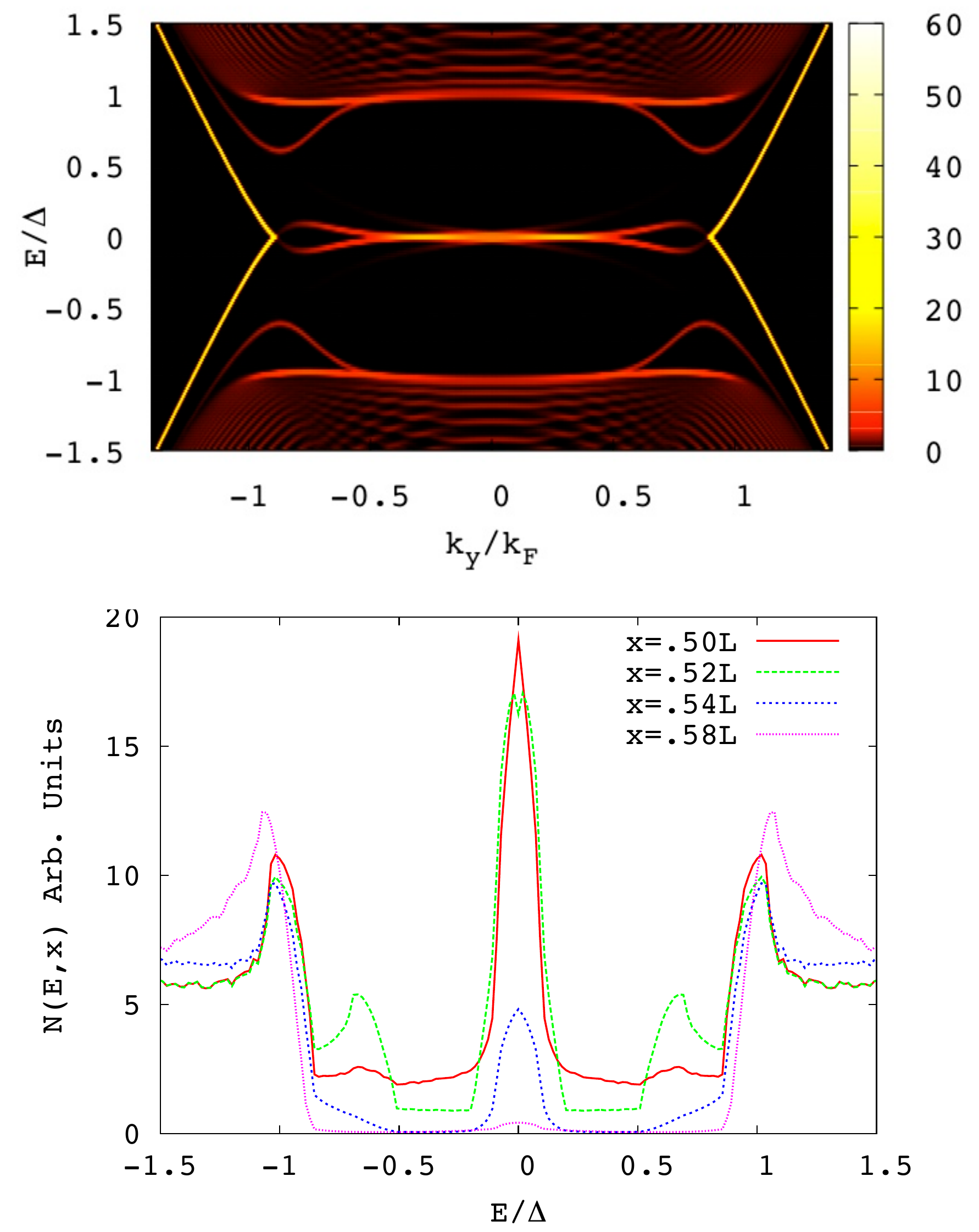}
\caption{(color online) The local spectral function $A_\uparrow(E,k_y,x)$
(upper panel) and local density of states $N(E,x)$ (lower panel,
red solid line) at the center of the junction,
$x=0.5L$. One sees ``flat" Andreev bound states near
zero energy for $-k_F<k_y<k_F$, and correspondingly a pronounced peak
at zero energy in the LDOS in the lower panel.
The lower panel also shows different LDOS away from the center,
for $x$ from $0.52L$ to $0.58L$. } \label{dos}
\end{figure}

The upper panel of Fig. \ref{dos} shows the spectral function
at the center of the junction, $A_\uparrow(E,k_y,x=0.5L)$ ($A_\downarrow$ is the
same for this value of $x$), with $\mu$=20meV,
$\Delta=5.5$meV,  $w=0.04L$, $L= 2576$nm, 
$ \hbar v_F$=4.1 \AA eV, and the Fermi momentum $k_{F}=\mu/(\hbar v_F)$.
In contrast to the $E\sim \hbar v_F k_y$ mode for $\mu=0$, we see 
Andreev bound states (ABS) near zero energy within a wide region $-k_F<k_y<k_F$,
where the slope $\hbar v_y=\partial E / \partial k_y$ approaches zero. 
The appearance of numerous crossings at exact zero energy for finite $k_y$ 
also agrees with the model calculation above in Fig.~\ref{jjsetup}d). 
Beyond this range, e.g. for $k_y>k_F$, the spectrum is reminiscent 
of the particle-hole folded dispersion of the helical metal, 
$E\sim \pm \hbar v_F(k_y- k_F)$.

As an approximate ansatz to describe the almost flat dispersion, we introduce the 
following phenomenological model for the ABS for large $\mu\gg \Delta$, 
\begin{equation}
E/\Delta = c (k/k_F)^N,
\label{bigN}
\end{equation}
where $c$ is a constant and $N$ is a large number. To fix $N$, we demand that
the slope of the dispersion at energy $E\sim \Delta$ coincides with that of the bare
dispersion, i.e., $\partial E/\partial k_y|_{E=\Delta}=\hbar v_F$. This gives an estimate
of $N$,
\begin{equation}
N\simeq \mu/\Delta.
\end{equation}
Note that we are only concerned with the ABS dispersion near zero energy and its continuation beyond $k_F$. 
For wider junctions, additional subgap ABS appear at finite energies,
and they are not described by Eq. (\ref{bigN}). Our ansatz is inspired by the 
mathematical theory of Dirac points with multiple topological charge $N$ as found in multilayered
system discussed in Ref. \cite{flat-N}.

The flat dispersion implies a peak at zero energy in the local density of states. The lower panel of 
Fig. \ref{dos} shows the LDOS at the center of the junction, at the S-TI boundary, and slightly into
the superconductor for the same junction parameters given above. 
While the zero energy peak becomes less pronounced when away from the junction center, 
it remains clearly visible and persists even into the superconductor. Thus, the predicted flat ABS has a clear
experimental signature in the tunneling conductance measurements.

The existence of two regimes including the flat Andreev bound states near zero energy is a general
feature. We have carried out systematic, self-consistent simulations for the 
general case of an inhomogeneous chemical potential, e.g., $\mu(x)=\mu_{TI}$ 
within the TI region and $\mu(x)=\mu_S\neq \mu_{TI}$ inside the superconductors.
For example, we fixed $\mu_S=100$meV, $\Delta$=10meV, $L=644$nm  and gradually increased $\mu_{TI}$ from zero to $\mu_S$ to monitor the evolution of the local spectrum with respect to the Fermi wave mismatch. 
We see the linear Majorana mode gradually changing into the flat ABS.

\begin{figure}[h]
\includegraphics[width=2.5in]{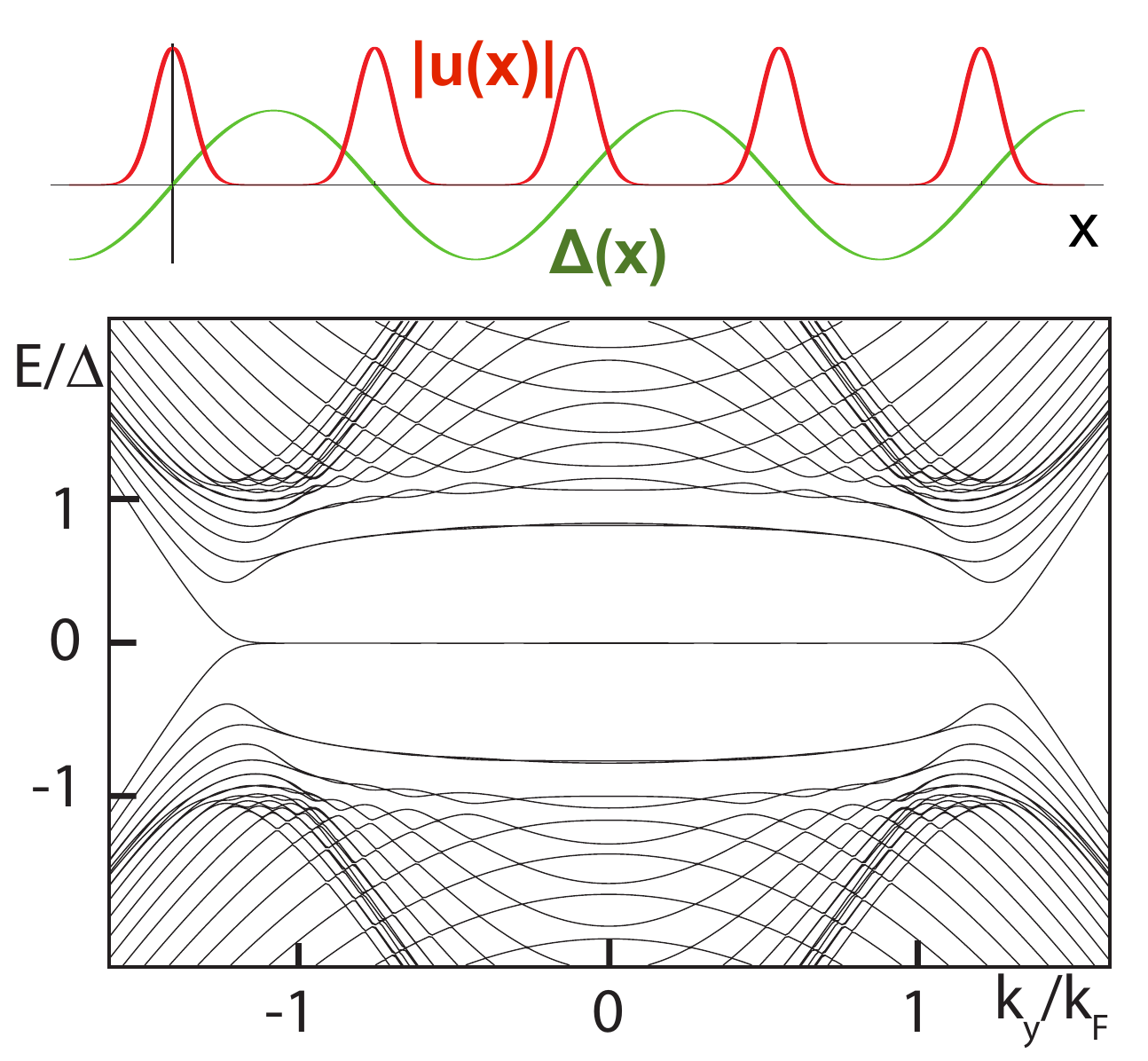}
\caption{(color online) Upper panel: Schematic of the periodic proximity structure with 
$\Delta(x)=\Delta  \sin(\pi x/a)$. The wave function $|u(x)|$ for the zero energy states
are peaked at the domain wall boundaries, $x=ma$. Lower panel: Energy spectrum for 
$a=24\hbar v_F/\Delta$ and $\mu=4\Delta$ is flat at zero energy, which has fine structures upon closer inspection.
}\label{array}
\end{figure}

Having established the existence of nearly flat ABS around zero energy, now
we systematically trace the evolution from the infinitesmal $\mu$, linear dispersing (Majorana) regime
to the large $\mu$ flat ABS regime. Also we would like to understand the details of ABS 
within its narrow ``band width". To this end, we will consider a simple model which generalizes the $\pi$ Josephson
junction to periodic systems. Namely, in Eq. (\ref{ham}), the order parameter modulates sinusoidally 
in the $x$-direction with period $2a$ as schematically shown in the upper panel of Fig. \ref{array}, 
\begin{equation}
\Delta(x)=\Delta  \sin(\pi x/a).
\end{equation}
The sign of the order parameter alternates. Thus the structure is effectively a periodic array of
the $\pi$ junctions discussed above in the limit $w \rightarrow 0$. One also recognizes that
$\Delta(x)$ describes a stripe/pair density wave, or Larkin-Ovchinnikov superconductor \cite{LO}.
While such superconductors are hard to find, one may imagine bringing them in contact with a TI to realize
the model consider here. Now the Hamiltonian $\mathcal{H}$ has discrete
translational symmetry in the $x$-direction, $\mathcal{H}(x)=\mathcal{H}(x+2a)$. We can apply the 
Bloch-Floquet theorem and introduce quasi-momentum $k_x$ living in the Brillouin zone of $(-\pi/2a,\pi/2a)$. 
For the prescribed $\Delta(x)$, the energy spectrum $E(k_x,k_y)$
can be obtained by diagonalizing $\mathcal{H}$ in $k$-space. Note that the TI (non-superconducting) region is shrunk
to a point, only the homogenous $\mu$ is left as tuning parameter.

\begin{figure}[h]
\includegraphics[width=3.4in]{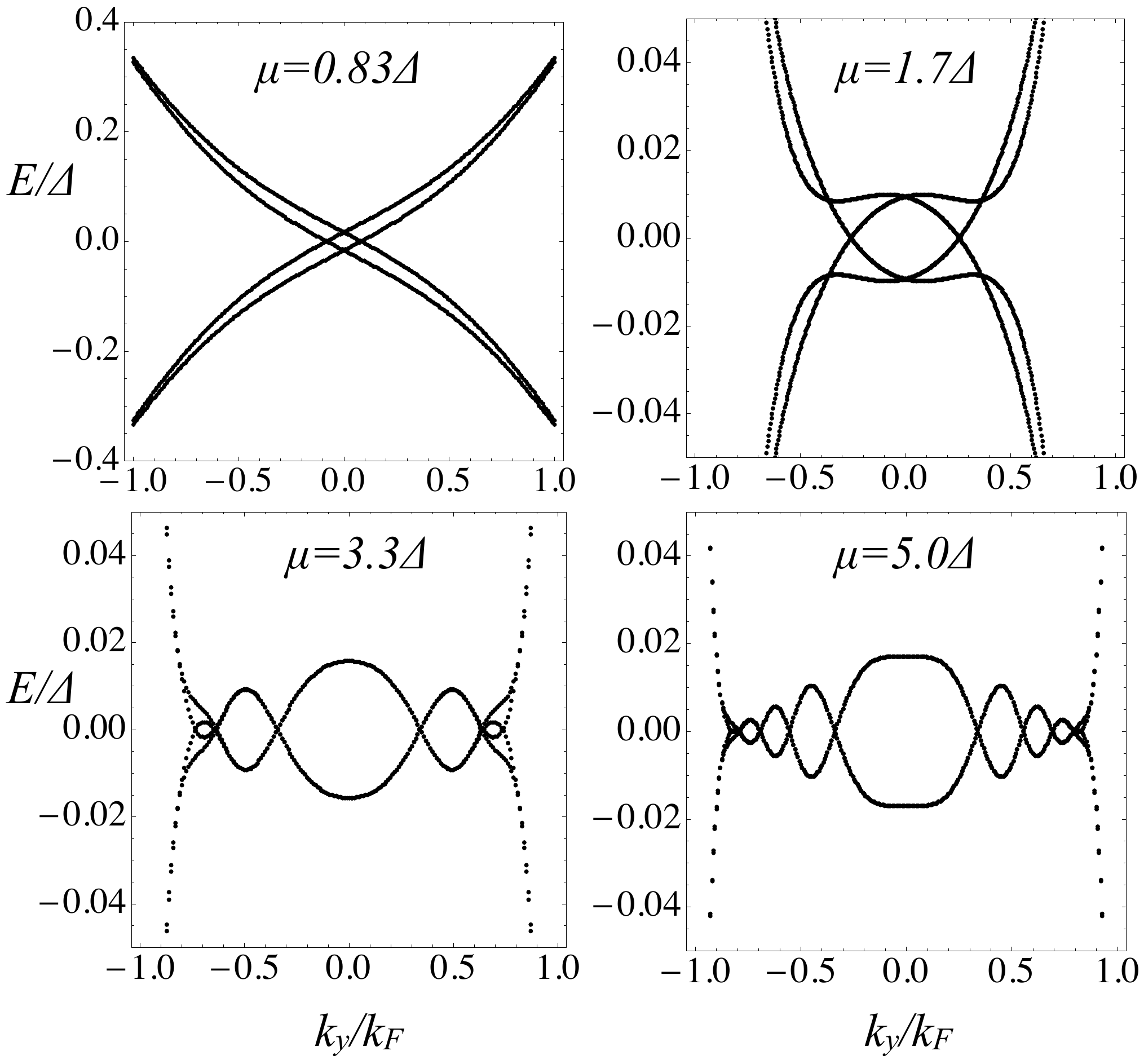}
\caption{Fine structures in the energy spectrum of the periodic proximity structure with fixed $a=12
\hbar v_F/\Delta$
and increasing $\mu$. The linearly dispersing Majorana spectrum at $\mu=0$ splits and develops
curvature to eventually become nearly flat within $(-k_F,k_F)$. The number of zero energy
crossings increases with $\mu$.
}\label{comp}
\end{figure}

The lower panel of Fig. \ref{array} shows the spectrum $E(k_x=0,k_y)$ for $a=24\hbar v_F/\Delta$, $\mu=4\Delta$. 
These flat ABS at zero energy do not show significant
variation with $k_x$. We have checked that the wave function of these zero energy states
are localized at the domain wall boundaries of the order parameter field, i.e., at $x=ma$ (red
curve in the upper panel of Fig. \ref{array}). For example, the wave function of the $k_y=0,k_x=0,E\approx 0$ 
mode can be fit well with periodic Gaussians $|u(x)|\propto \exp(-{1.85 (\pi x/\sqrt{2}a)^2})$.
Since $a$ is large in this case, these results agree well with the single junction result before.
The dispersion, for example, can be fit well using the ansatz in Eq. (\ref{bigN}).
The vanishing band width is, of course, only valid on coarse scales. Closer inspection, by blowing up the spectrum near zero as
illustrated in Fig. \ref{comp}, reveals the busy life of the ABS with $N_c$ crossings at zero energy,
where $N_c$ scales linearly with $\mu$, in agreement with Fig. \ref{jjsetup}d).
Remarkably, all these fine details are compressed within a small energy range. 

Fig. \ref{comp} illustrates the evolution of the ABS at low energies for the periodic structure
as $\mu$ is increased from zero. For small value of $\mu=0.83 \Delta$, the linear Majorana dispersion splits
into two, each developing a curvature, as
the zero energy crossings move to finite $k_y$ values. Further increasing $\mu$,
these two crossings are stretched further outwards, while the dispesion within $k_y\in(-k_F,k_F)$ 
begin being bent and stretched to form the precursor of the flat band. At the same time,
addition of new crossings introduces more twists. The number of crossing scales with $N_c\sim \mu/\Delta$.
The spaghetti now becomes a rope, and looking from
afar, it appears as a thin thread.

Flat bands are more novelties than the norm in condensed matter \cite{heikkila_flat_2011}. Recently, several authors have
demonstrated that {\it surface} Andreev bound states with flat dispersion arise in certain topological 
superconductors, for example Cu$_x$Bi$_2$Se$_3$ \cite{hsieh_majorana_2012} and 
non-centrosymmetric superconductors \cite{PhysRevB.84.020501, schnyder_topological_2011}. 
In addition, flat bands within a superconducting gap could precipitate secondary interactions between quasiparticles. 
Their existence can be traced
back to the nontrivial topology associated with the gapped bulk, and thus are topologically protected. 
This mechanism giving rise to flat bands, via the bulk-boundary correspondence,
differs from what is considered here. For example, in Ref. \cite{hsieh_majorana_2012},
a robust crossing at $k=0$ is a crucial point in the argument, and the total number of zero energy 
crossings is guaranteed an odd number. In our case, states at $k_y=0$ are gapped
for finite size systems (or finite period $2a$). Despite these differences, the
zero modes share the common trait that they are associated with the sign change of the order parameter
when electrons are reflected at the surface or interface.

Several groups have successfully fabricated Josephson structures on Bi$_2$Se$_3$ of various
length using a variety of 
superconducting materials including Al, Al/Ti, W, Nb, and Pb etc. \cite{SacACpAC:2011vn,
PhysRevB.84.165120,
Veldhorst:2012uq,Qu:2012kx,2012arXiv1202.2323W}. 
Gate tunable supercurrent
has been observed and argued to be due to the TI surface state \cite{SacACpAC:2011vn}. Superconducting quantum 
interference devices based on such junctions have also been demonstrated \cite{2011arXiv1112.5858V,Qu:2012kx}. 
Thus the flat
Andreev bound states at zero energy, and the zero bias conductance peak in the local 
density of states, predicted here should be experimentally accessible. 
Future work will explore control of these slowly dispersing Andreev levels
working as qubits \cite{PhysRevLett.90.087003}
when confinement in the $y$ direction is also introduced. Our work
also suggests the ac dynamics of the S-TI-S junctions will likely to be very complex
featuring different regimes.
The flat ABS at zero energy predicted for periodic junction arrays may potentially find 
technological applications. For example, a diverging density of states at the midgap
may be used to generate microwave resonances.

We would like to thank Noah Bray-Ali, Liang Fu, and Takuya Kitagawa for helpful discussions.
This work is supported by ONR grant No. N00014-09-1-1025A.

\bibstyle{aip}
\bibliography{jj}{}

\begin{thebibliography}{24}
\bibitem[{foo()}]{footnote1}
\bibinfo{note}{The exact choice of boundary conditions will not affect the
  local physics at the junction, by choosing $L$ to be large. The open boundary
  condition may be realized, as an example, by having strong ferromagnets at
  $x=0,L$. They will intoduce chiral edge modes spatially well seperated from
  the center of the junction and will not affect the local spectrum there.}
\end{thebibliography}

\end{document}